\documentclass[a4paper,twoside]{article}
\newcommand{\affil}[1]{$^{\rm #1}$}
\usepackage[authoryear]{natbib}
\bibpunct{(}{)}{;}{a}{}{,}
\usepackage{graphicx}
\date{} 
\newcommand\aj{\rmfamily{AJ}}%
\newcommand\apj{\rmfamily{ApJ}}%
\newcommand\apjs{\rmfamily{ApJS}}%
\newcommand\apjl{\rmfamily{ApJL}}%
\newcommand\aap{\rmfamily{A\&A}}%
\newcommand\aaps{\rmfamily{A\&AS}}%
\newcommand\mnras{\rmfamily{MNRAS}}%
\newcommand\physrep{\rmfamily{Phys.~Rep.}}%
\newcommand\pasp{\rmfamily{PASP}}%
\title{\large\bf\flushleft Subtraction of Bright Point Sources from Synthesis Images of the Epoch of Reionization}
\author{\parbox{\textwidth}{\flushleft
\vspace{-0.5cm}
{\it B. Pindor\affil{A,D}}
{\it J.~S.~B.~Wyithe \affil{A}}
{\it D.~A.~Mitchell \affil{B}}
{\it S.~M.~Ord \affil{B}}
{\it R.~B.~Wayth \affil{C}}
{\it L.~J.~Greenhill \affil{B}}\\
\vspace{0.4cm}
{\small \affil{A}\,University of Melbourne, School of Physics, Parkville 3010, Australia}\\
{\small \affil{B}\,Harvard-Smithsonian Center for Astrophysics, 60 Garden Street, Cambridge, MA 02138-1516}\\
{\small \affil{C}\,ICRAR~/Curtin Institute of RadioAstronomy, GPO Box U1987, Perth, WA6845, Australia}\\ 
{\small \affil{D}\,Email: bpindor@unimelb.edu.au}}}

\begin{document}
\twocolumn[
\begin{changemargin}{.8cm}{.5cm}
\begin{minipage}{.9\textwidth}
\vspace{-1cm}
\maketitle
%
%
\small{\bf Abstract: Bright point sources associated with extragalactic AGN and radio
 galaxies are an important foreground for low frequency radio experiments
 aimed at detecting the redshifted 21cm emission from neutral hydrogen during the epoch
 of reionization. The frequency dependence of the synthesized beam implies that
 the sidelobes of these sources will move across the field of view as a
 function of observing frequency, hence frustrating line-of-sight
 foreground subtraction techniques. We describe a method for subtracting
 these point sources from dirty maps produced by an instrument such as
 the MWA. This technique combines matched filters with an iterative
 centroiding scheme to locate and characterize point sources in the
 presence of a diffuse background. Simulations show that this technique
 can improve the dynamic range of EOR maps by 2-3 orders of magnitude.}


\medskip
\medskip
\end{minipage}
\end{changemargin}
]
\small

\section{Introduction}

Highly redshifted 21cm radiation emitted by neutral hydrogen gas during the Epoch of Reionization (EOR) contains a wealth of cosmological and astrophysical information. Several experiments aimed at detecting and characterizing this radiation are currently in progress or under construction (MWA\footnote{http://www.mwatelescope.org/}, PAPER\footnote{http://astro.berkeley.edu/\textasciitilde dbacker/eor/}, LOFAR\footnote{http://www.lofar.org/}, GMRT\citep{2008AIPC.1035...75P}). The EOR signal is also one of the main science targets of next generation radio facilities such as the SKA\footnote{http://www.skatelescope.org/}.

Observations of the Gunn-Peterson trough in the spectra of high-redshift quasars indicate that reionization was essentially complete by redshift $z =6$ \citep{2001AJ....122.2850B}. Meanwhile, the optical depth due to Thompson scattering observed for CMB photons by the WMAP satellite implies that for an instantaneous process $z_{reion} = 10.9 \pm 1.4$ \citep{2009ApJS..180..330K}. Together, these observations support theoretical expectation that reionization was an extended process \citep{2009arXiv0908.3891P}.  The expected redshift range of reionization, $z \simeq 6-15$, puts the 21cm line at 200-90 MHz. Such low radio frequencies poses a number of observational challenges. Terrestrial transmissions from radio, television and satellite communications are all prominent at or near this band. Additionally, refraction of low frequency radio waves by the ionosphere introduces time-variable distortions into the observed radio sky which require continuous re-calibration \citep{2008ISTSP...2..707M}. 

Observations of the EOR signal are further complicated by astrophysical foregrounds. These foregrounds are dominated by galactic diffuse synchrotron emission (GDSE) and extragalactic AGN (point sources), with a smaller contribution from galactic free-free emission  \citep{1999A&A...345..380S}. The GDSE has a steep spectral index \citep[$\beta \sim -2.5$,][]{2008AJ....136..641R} which makes it hundreds of times brighter at 150MHz than in the 21cm rest-frame, while extragalactic point sources have a typical spectral index of $\beta \sim -0.8$ \citep{2002IAUS..199...58S}, with some evidence of flattening at lower frequencies \citep{2004ApJS..150..417C}. Interferometric observations are sensitive only to fluctuations in the sky brightness, and, taken together, these foregrounds are expected to have fluctuations at least three orders of magnitude greater than the EOR fluctuations \citep{2005ApJ...625..575S}. Initial observations of the diffuse foregrounds confirm these expectations \citep{2009A&A...500..965B,2008MNRAS.385.2166A,2010arXiv1002.4177B}.

Fortunately, the daunting task of isolating the comparatively faint EOR signal is made tractable by the spectral smoothness of the foregrounds. The GDSE exhibits an essentially featureless power-law spectrum, while the combination of many point sources of varying spectral indices can also be well-reproduced as a smoothly-varying function of frequency \citep{2006ApJ...650..529W}. In contrast, observing the EOR signal at different frequencies corresponds to probing the relatively rapidly varying IGM density field across a substantial range of redshifts. Numerous authors have taken advantage of this  distinction to demonstrate the possibility of subtracting foregrounds by fitting polynomials or other smoothly-varying functions along the observing frequency / line-of-sight dimension \citep{2006ApJ...650..529W,2009MNRAS.397.1138H,2009ApJ...695..183B,2008MNRAS.391..383G,2008MNRAS.390.1496G}.

The effectiveness of these foreground removal strategies assumes the prior removal of the brightest point sources. Bright point sources cannot be effectively removed through line-of-sight subtraction because the necessarily incomplete $uv$-coverage of any real interferometer inevitably creates a sidelobe pattern, evocatively termed "frizz" by \citet{2009MNRAS.394.1575L}, across the plane of the sky. Changing the observing frequency changes the size of the synthesized beam and consequently moves this "frizz" across any given point on the sky. In this way, high angular-frequency structure in the sidelobe pattern across the plane of the sky enters the observational frequency dimension, causing "mode-mixing" in the line-of-sight power spectrum \citep{2009ApJ...695..183B}. For this reason, most previous foreground removal studies assume there is some flux threshold $S_{cut}$ above which all point sources can be removed. \citet{2009ApJ...695..183B} and \citet{2009MNRAS.398..401L} agree that $S_{cut}$ should be of order $10-100$ mJy. However, the method by which these bright sources are to be removed remains an open question. In this paper, we introduce a new technique for subtracting bright point sources from synthesis images produced by an instrument such as the MWA.  

\section{The MWA EOR Experiment}

The primary goal of the MWA EOR experiment is to measure the power spectrum of the cosmic neutral hydrogen density field. 21cm radiation from neutral hydrogen will be visible against the cosmic microwave background (CMB) when the spin temperature, $T_S$, deviates from the CMB temperature, $T_{CMB}$ \citep{1952AJ.....57R..31W,1959ApJ...129..536F} with a predicted differential brightness temperature given by \citep{2003ApJ...596....1C}:

\begin{eqnarray}
\delta T_b = 26 \mbox{ mK } x_{HI} (1 + \delta) \left (1 - \frac{T_{CMB}}{T_S} \right) \left(\frac{\Omega_{b} h^2}{0.02} \right) \nonumber \\ 
\left [ \left ( \frac{1+z}{10} \right) \left ( \frac{0.3}{\Omega_m} \right) \right] ^{1/2}
\end{eqnarray}

\noindent During reionization, $T_S > T_{CMB}$, and hence neutral hydrogen can be detected in emission. 

The detectability of the cosmological 21cm signal is fundamentally limited by the thermal noise of the radio telescope. For an interferometer, each visibility is subject to a thermal noise contribution given by  \citep[eg][]{2001isra.book.....T}

\begin{eqnarray}
\label{noise_eqn}
V_{rms} = \frac{2 k_B T_{sys}}{A_e \sqrt{df \tau}},
\end{eqnarray}

\noindent where $T_{sys}$ is the is the system temperature, $A_e$ is the effective area of a single antenna, $df$ is the channel bandwidth, and $\tau$ is the integration time. For the MWA, the system temperature is dominated by the sky and is expected to be $\sim$ 180K at 178 MHz \citep{2006PhR...433..181F}. For an isotropic power spectrum\footnote{The neutral hydrogen density field is expected to undergo some cosmological evolution over the full 32 MHz MWA bandpass, but this does not affect the present discussion}, this thermal noise is averaged over the thousands of baselines which sample each three-dimensional Fourier mode, leading to a formally highly significant detection.

The ability of the MWA to detect the EOR signal is in practice limited by the accuracy of the instrument calibration and foreground subtraction. As described in \citet{2008ISTSP...2..707M}, the high data rate output by the MWA correlator precludes the long-term storage of the raw visibilities, necessitating real-time calibration and imaging. The calibration cadence is set by the need to i) resolve temporal variation in the ionospheric refraction and ii) update the direction-dependent antenna response model as sources move across the sky. The MWA will complete a calibration and imaging cycle every 8 seconds. The calibrated images formed from these 8s integrations will then be coadded into stacks of $\sim 8$ minutes of observing time to further reduce the data storage requirements. These co-added 8 minute dirty maps form the basic data product from which the offline data analysis of the EOR experiment will be done.

\section{A Description of the Method}

\subsection{Motivation}

Before describing our subtraction technique, we briefly review the treatment of point sources in a number of observational regimes which will motivate our approach. 

\subsubsection{Radio Astronomy}

The most well known and widely practiced method for removing point sources in radio astronomy is the CLEAN algorithm \citep{1974A&AS...15..417H} and its many variants. CLEANing addresses one of the central difficulties in processing synthesis images; that it intrinsically involves deconvolution. The Cotton-Schawb CLEAN variant \citep{1984AJ.....89.1076S}, which subtracts the clean components from the ungridded visibilities and hence allows one to avoid aliasing and gridding errors, generally produces the best results. Nonetheless, the obtained dynamic range is usually limited to $\sim 10^3$ \citep{1992ASPC...25..170B}. More sophisticated approaches have been able to achieve dynamic ranges of  $\sim 10^6$ \citep{2008A&A...490..455C, 2004ExA....18...13V}, however these techniques require real-time processing which would almost certainly exceed the MWA computing resources. Additionally, despite some theoretical efforts in quantifying the uncertainties associated with CLEANed images \citep{1978A&A....65..345S}, practical experience has shown that CLEANing often alters image statistics and leaves spatially correlated residuals \citep[ie 'stripes',][]{1999ASPC..180..151C} which would likely corrupt measurements of the EOR signal. 

A second approach to subtracting radio point sources is 'Peeling' \citep{2004SPIE.5489..817N,2007ITSP...55.4497V,2009A&A...501.1185I}. Peeling is essentially the sequential self-calibration and subtraction of the brightest sources in the field and is intended to overcome the calibration difficulties introduced by ionospheric refraction and direction-dependant gains which are inevitable in wide-field low frequency radio observations. Peeling has the advantage of subtracting the point source model from the ungridded visibilities, similar to the Cotton-Schawb CLEAN variant. However, due to the need to update the calibration solution on timescales which are short compared to the timescale for ionospheric variations, it is unlikely that more than $\sim 100$ sources will be peeled by the MWA Real-Time Calibration, leaving hundreds of bright sources above the confusion limit. Additionally, the expected data rate from the MWA correlator ($\sim$ 19 Gb/s) precludes long-term storage of the raw visibilities, implying that post-processing to remove foregrounds will likely be restricted to the gridded visibilities / dirty maps. 

\subsubsection{CMB}

Removal of point sources is also an important foreground subtraction step in processing of Cosmic Microwave Background (CMB) temperature maps. Matched filters are used to increase the contrast between the point sources and the CMB or other diffuse components \citep{1998ApJ...500L..83T}. CMB beams are sufficiently compact that it suffices to identify and mask out the bright pixels associated with the main lobe. 

\subsubsection{Optical Astronomy}

Locating and measuring the flux of point sources are the most fundamental processes in optical astrometry and photometry. Images are usually convolved with the Point Spread Function (PSF) to maximize signal-to-noise. For reasonably oversampled images, it is routine to obtain astrometric centroids on the order of 10\% of the pixel scale. \citep{2003AJ....125.1559P}.

\subsection{Subtraction Procedure}
\label{Subtraction Procedure}

Based upon the above considerations, we designed a procedure for subtracting bright point sources from synthesized radio images produced by an interferometric instrument such as the MWA. As mentioned above, the very high MWA data rate and a finite storage capacity imply that only time averaged dirty maps will be available for offline data analysis. Given a dirty map, our subtraction proceeds as follows:

i) \textbf{Filtering:} We use a matched filter to optimize the signal-to-noise ratio and reduce contamination in the flux and centroid estimates caused by diffuse sky emission (GDSE). As shown by \citet{1996MNRAS.279..545H}, the optimal linear filter, $\psi$, is one which upweights the source profile, while downweighting the scales at which the generalized noise (ie all non-pointlike components in the sky) is prominent. In Fourier space, this implies

\begin{eqnarray}
\hat{\psi}(\mathbf{k}) \propto \hat{\tau}(\mathbf{k}) / P(\mathbf{k})
\label{MF_equation}
\end{eqnarray}

We approximate $\tau$, the source profile, as a two-dimensional Gaussian fitted to the main lobe of the synthesized beam at the centre of the FOV. $P(\mathbf{k})$ is the power spectrum of the generalized noise. If there is no diffuse sky component, then $P(\mathbf{k}) = 1$ and matched filtering is equivalent to convolving the dirty image with our Gaussian beam model. If there is a diffuse sky component, then we estimate $P(\mathbf{k})$ iteratively from the data. Specifically, we first carry out our subtraction procedure assuming that $P(\mathbf{k})$ is constant. The result of this initial subtraction is a residual image with most of the point source flux removed. We then measure $P(\mathbf{k})$ from this residual image and fit the measured values to a second order polynomial in log space (ie $ \log(P(k)) = a (\log(k))^2 + b (\log(k)) + c$). We then repeat our subtraction procedure using this model of $P(\mathbf{k})$ in Equation \ref{MF_equation}. In our simulations, we proceed to identify point sources as $5\sigma$ local maxima in the filtered images. For future real data, it is likely that the locations of most bright sources would be known \textit{a priori} from higher frequency observations. 

ii) \textbf{Centroiding:} We use the centroiding procedure developed for the Sloan Digital Sky Survey (SDSS, \citep{2000AJ....120.1579Y}) photometric pipeline to measure the positions of detected sources in the filtered maps. This procedure uses Gaussian quartic interpolation to predict the centroid based on a local maximum and the eight surrounding pixels. There are two subtleties in this centroiding procedure: i) the main lobes are not actually Gaussian, or even symmetric. Hence, the measured centroids are biased and the bias is direction dependent.  We correct for this bias in the same manner as the SDSS; by simulating the beam at the inferred centroid position and determining the offset between the expected and measured positions ii) centroid and flux measurements of any given source are perturbed by the sidelobes of the other bright sources in the field. We address this problem through iteration; an initial estimate is made of the position and flux of each source neglecting the contribution of other sources, and this estimate is subsequently used to correct for the sidelobe contribution at the position of every source. In principle, this process could be repeated until the parameters for each source converged, but in practice the generation of each source with sub-pixel positional accuracy is computationally intensive and the results presented in this work are based on a single such iteration.  

iii) \textbf{Aperture Photometry:} We estimate the flux of a source through the use of a circular aperture at the measured centroid position in the match filtered image. We correct for aperture bias by simultaneously measuring the aperture flux of a match filtered image of the beam reconstructed at the centroid position. 

\section{Simulations}

\subsection{Description of Simulations}
\label{simulations}

We developed and tested our subtraction method through the use a simulation pipeline which combines MAPS, the MIT Array Preformance Simulator (Wayth et al, in preparation), with the MWA calibration and imaging system (MWA RTS, Mitchell 2008). MAPS is software package for simulating the observed visibilities generated by a real interferometric array. In our case, it has been configured to simulate a 512 tile array with the antenna design and provisional layout of the MWA. Each simulation is a two second snapshot integration at an observational frequency of 178 MHz and over a channel bandwidth of 40 kHz. These simulations are computationally intensive as each integration contains over $10^5$ visibilites. Consequently, we make two important simplifications at this stage. 

First, we ignore the effects of thermal noise. Each visibility should be subjected to a thermal noise contribution given by equation \ref{noise_eqn}. However, this level of thermal noise is not appropriate for EOR foreground subtraction as the foregrounds will be subtracted not from snapshots but rather from coadded maps which result from hundreds of hours of integration. Naively, we could introduce a longer integration time $\tau$ to scale the thermal noise, but this does not account for the scaling of the sidelobe level which would result from the simultaneous rotation synthesis. The correct scaling could be obtained by generating and coadding snapshots corresponding to a $\sim$ 6 hr exposure, however such a simulation is beyond our present computational resources and is a subject for future work. 

Second, we assume that all MWA tiles have identical response (gain). By extension, this assumption implies that the MWA beam in our simulations is direction-independant. On the other hand, our imaging procedure and point source subtraction technique do not make this assumption. Rather, they treat the data as in the general case of a direction-dependant beam. In future we intend to simulate a representative variation of tile gains, but in this work we only consider the simplest case of identical tiles. MAPS also has the ability to simulate the effect of the ionosphere on the observed visibilities for a given model sky, but in this work we neglect this effect. 

\begin{figure}[h]
\begin{center}
\includegraphics[scale=0.3, angle=0]{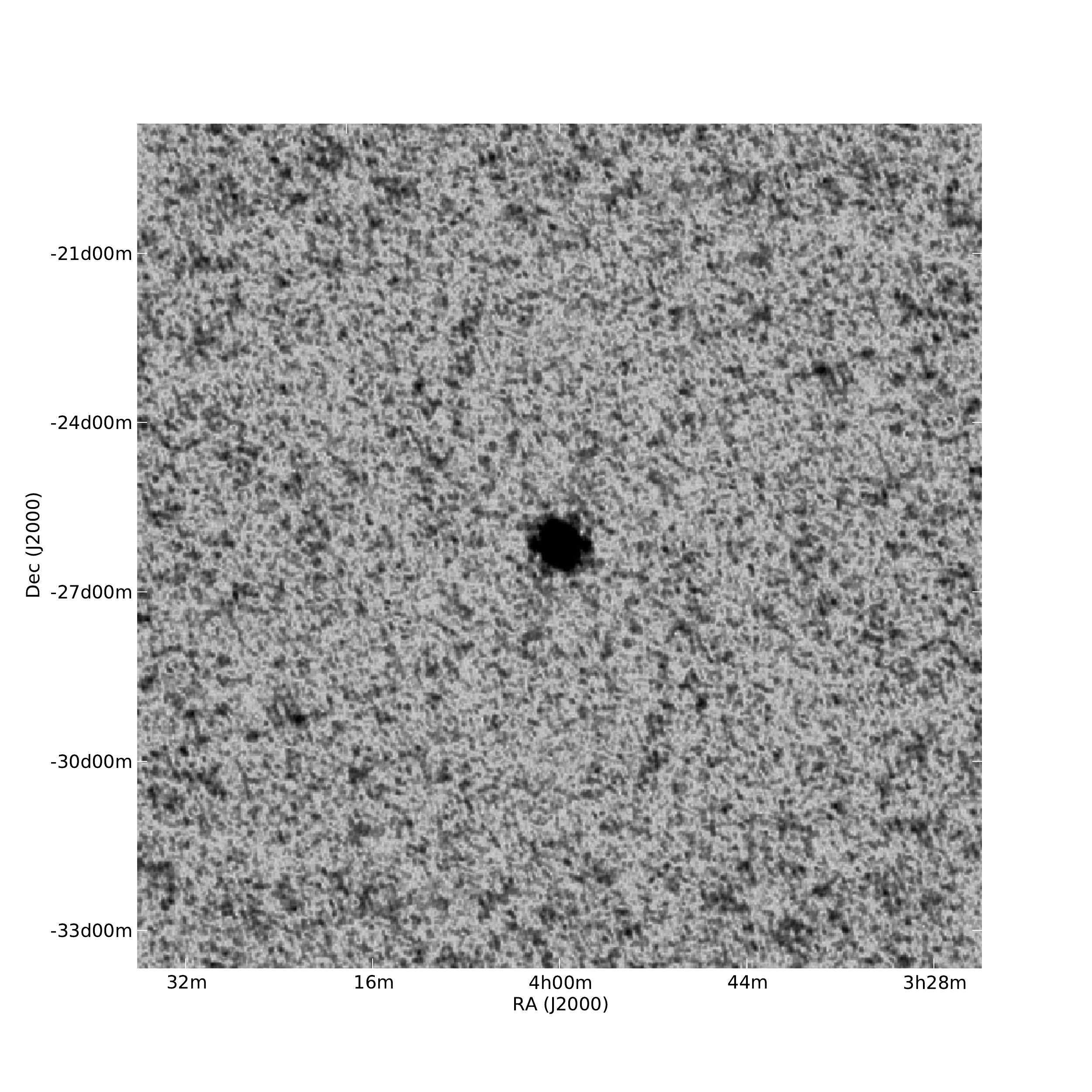}
\caption{The synthesized beam for a 512 element MWA. A logarithmic stretch has been applied in order to emphasize the extended sidelobe structure.}
\label{beamFig}
\end{center}
\end{figure}

The simulated visibilities produced by MAPS are subsequently processed into synthesized images by the MWA RTS. The resulting SIN projection images have a pixel scale of 1.9 arcmin/pixel at the field centre. We cropped the central 512 by 512 pixels of each image (corresponding to a FOV of $\sim$ 18 by 16 deg) and restrict our analysis to this region. This same simulation procedure is also used to reconstruct the synthesized beam at any point in the field. The resulting synthesized beam at the field centre is illustrated in Figure \ref{beamFig}. The sidelobe fluctuations seen throughout the field are $\sim 1$ \% of the peak source flux.

\section{Simple Tests}

\subsection{Single Point Sources}

We begin by simulating 100 realizations of a randomly placed, isolated point source in an empty sky to measure the accuracy of our centroiding algorithm in the ideal case. These point sources have a flux of 100 Jy/beam, although the flux normalization is not meaningful in this simple case. Applying our centroiding algorithm directly to the dirty maps yields an RMS centroiding error of 0.02 pixels. Measuring the centroid in the filtered images ($P(\mathbf{k}) = 1$ in this case) yields an RMS centroiding error of 0.002 pixels. Following correction for the centroiding bias described in Section \ref{Subtraction Procedure}, we obtain an RMS centroiding error of $10^{-4}$ pixels. These errors and all subsequent results are summarized in Table \ref{centroiding table}.

\begin{figure}[h]
\begin{center}
\includegraphics[scale=0.4, angle=0]{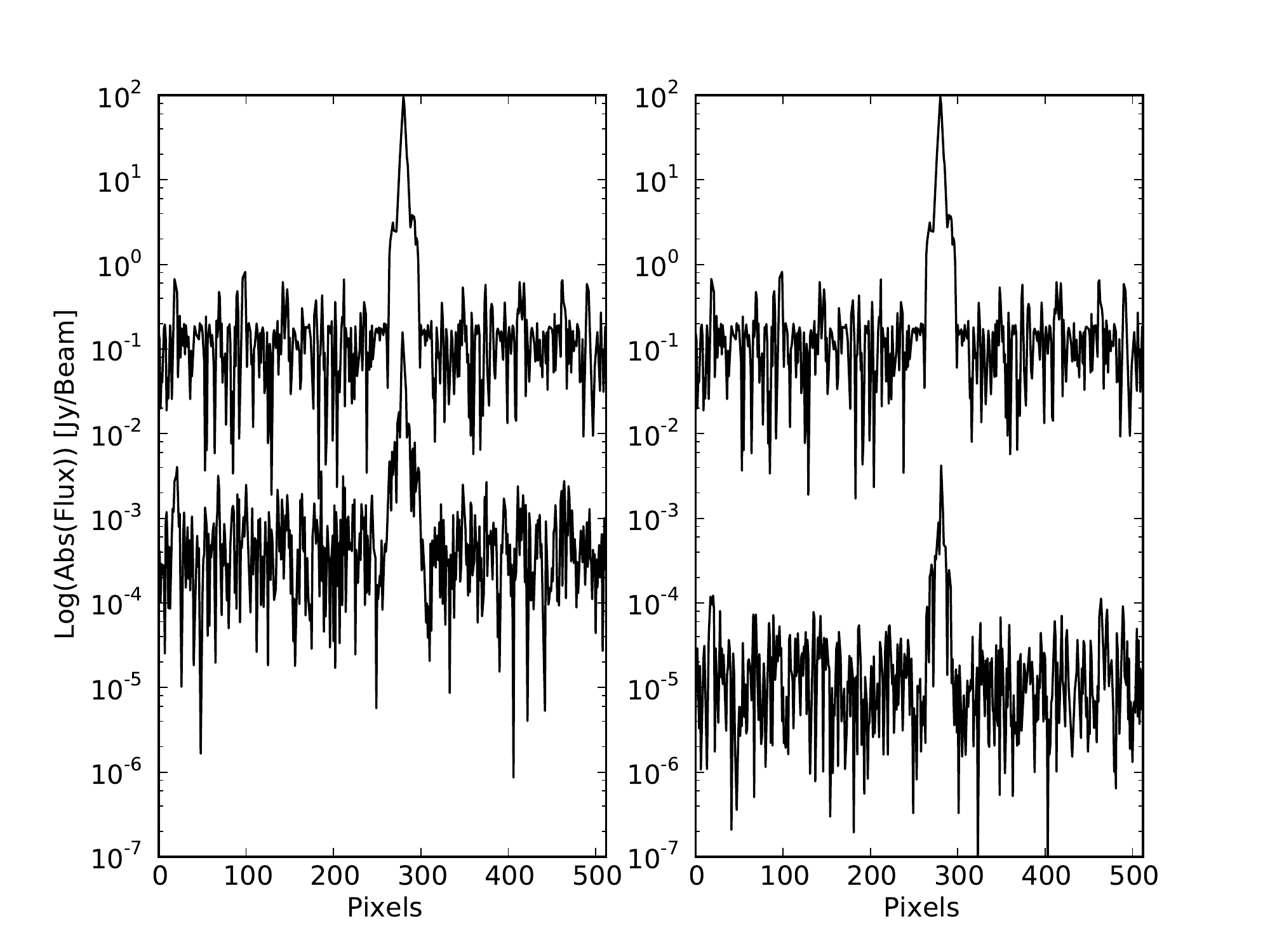}
\caption{The results of point source subtraction for a single source. In the left panel, the upper line represents the raw image and the lower line represent the residuals following subtraction from the Match Filtered image. In the right panel, the upper line represents the raw image and the lower line represent the residuals following subtraction from the Match Filtered image with the PSF bias correction applied. See text for more details.}
\label{residuals_2panel}
\end{center}
\end{figure}

We also measure the change in the dynamic range of our simulated images following point source subtraction. The dynamic range is conventionally defined as the peak brightness divided by the RMS in empty parts of the image. We define empty regions as those pixels with counts less than twice the maximum value observed in our diffuse sky model (see below). In practice, this threshold is about 3\% of the maximum source flux.  We are most interested in the \textit{relative} dynamic range; the ratio of the dynamic range between the raw and subtracted images. Table \ref{centroiding table} lists the dynamic range for the raw (unsubtracted) images, the Matched Filter subtracted images, and the Matched Filter and PSF bias corrected subtracted images. The listed values are the logarithmic means of our 100 realizations. The relative dynamic range for the Match Filtered and PSF bias corrected subtracted images is $2.0\times 10^4$. We illustrate the subtraction results for a single simulated source in Figure \ref{residuals_2panel}. Both panels in this figure show cross-sections through the brightest pixel in the field. The upper line in both panels is the raw image. We have plotted the absolute values to allow for a logarithmic scale. The point source is clearly visible near the center of the field, as are the $\sim 1\%$ sidelobes. In the left panel, the lower line shows the residuals following subtraction from the Match Filtered image. The relative dynamic range in this case is $\sim 240$. Note that the subtraction for this particular source happens to be substantially less accurate than the average of 100 realizations reported in Table \ref{centroiding table} (ie $7.9\times10^2$). In the right panel, the lower line shows the residuals following subtraction from the Match Filtered image where the centroid has been corrected for PSF bias. The relative dynamic range in this case is $\sim 7.6 \times 10^{3}$.  In both cases, significant residuals remain at the position of the point source. The question of whether these pixels should be masked out or if they will be further reduced by continuum foreground subtraction is a subject for future work.  

\subsection{Single Point Sources with a Diffuse Background}

\begin{figure}[h]
\begin{center}
\includegraphics[scale=0.3, angle=0]{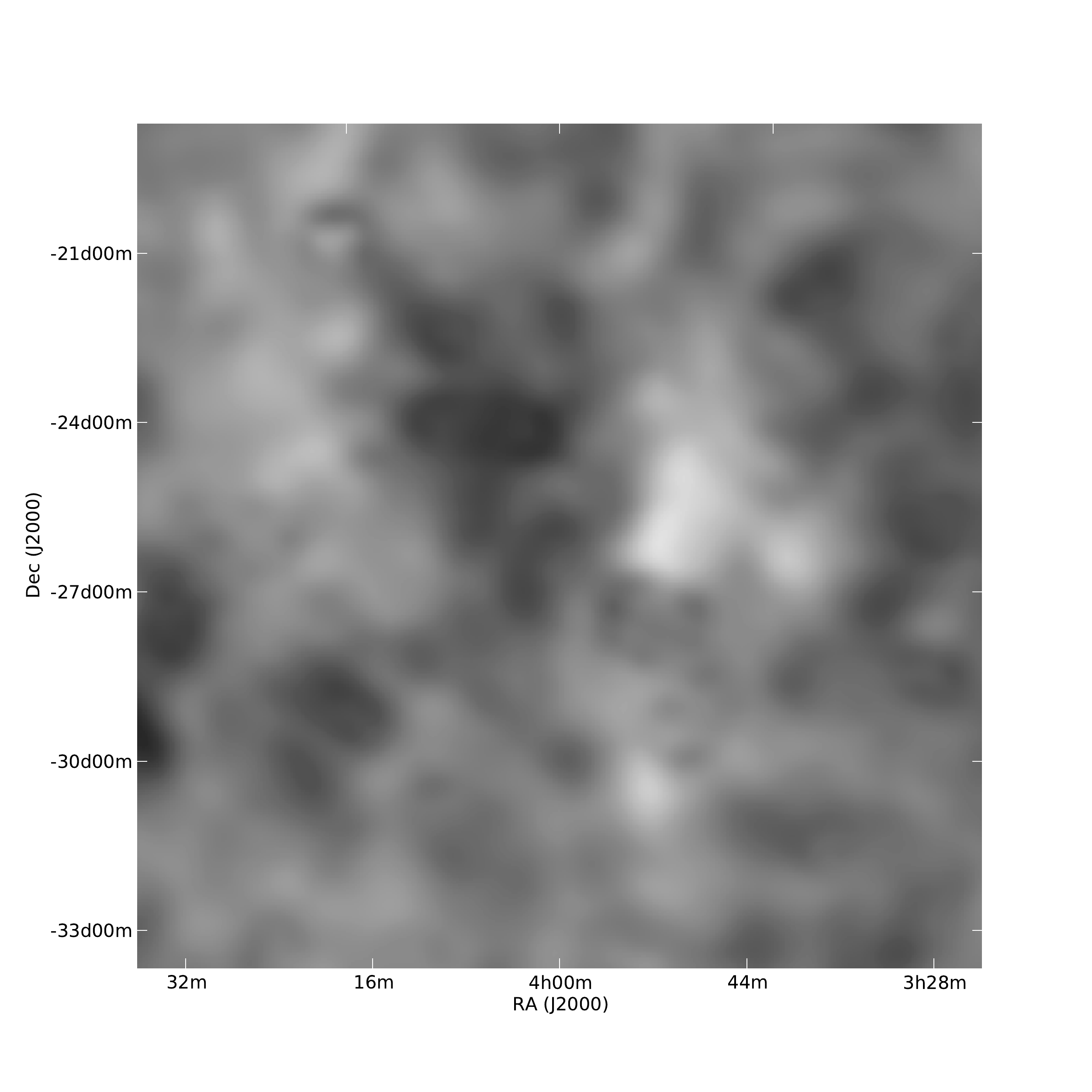}
\caption{The diffuse sky component generated from de Oliveira-Costa et al. (2008). A linear stretch has been applied in this case.}
\label{skyFig}
\end{center}
\end{figure}

Next, we add a diffuse sky component based upon \citet{2008MNRAS.388..247D}. We use a scaling appropriate to an observational frequency of 178 MHz, and locate the center of our field  at the planned center of the main MWA EOR field (RA 4h, Dec -26$^{\circ}$). Hence, this diffuse component is identical for all of our simulations.  Figure \ref{skyFig} illustrates the resulting diffuse emission. As before, we simulate 100 realizations of a randomly placed, isolated point source which are added to this diffuse background. The Matched Filter in this case uses an empirical $P(\mathbf{k})$ as described in Section \ref{Subtraction Procedure}. The RMS centroiding error following Matched Filtering and PSF bias correction is $5.5\times 10^{-4}$ pixels. 
In order to be able to directly compare the results of these simulations to those performed with no diffuse component, we subtracted a simulated image of the diffuse sky without any point sources (ie the result of a perfect point source subtraction) from the residual image prior to calculating the dynamic range, so that the reported value is the dynamic range of the point source residuals only. To clarify, this subtraction of the diffuse sky is performed only after we have completed our point source subtraction procedure with the diffuse component present. The relative dynamic range for the Match Filtered and PSF bias corrected subtracted images is $2.5\times 10^3$.

\subsection{Multiple Point Sources}
\label{10source_emptySky}

The centroid and flux estimates for each source are perturbed by the presence of all other sources in the field. We investigated this effect by repeating our previous simulations with 10 randomly placed sources of equal (100 Jy/beam) flux in an empty sky. Equal flux sources correspond to the case of maximal mutual perturbation. Each realization has a different random distribution of sources within the FOV. The iterative correction for source centroids and fluxes described in Section \ref{Subtraction Procedure} was applied. The RMS centroiding error following Matched Filtering and PSF bias correction is $5.9\times 10^{-4}$ pixels and the corresponding improvement in the dynamic range is $1.8\times 10^3$. We note that the raw dynamic range in this case is predictably lower than in the single source case due to the increased sidelobe noise.

\subsection{Multiple Point Sources with a Diffuse Background}

We added our diffuse sky model to the realizations of 10 randomly placed equal flux point sources. The RMS centroiding error following Matched Filtering and PSF bias correction is $5.5\times 10^{-4}$ pixels and the corresponding improvement in the dynamic range is $1.9\times 10^3$. Notably, the results are effectively the same with and without the diffuse component for these sufficiently bright sources.

\subsection{Comparison with Peeling}
\label{peeling_section}

We compared the results of our subtraction technique to the effectiveness of the MWA RTS peeling routine. As described above, peeling is performed on the ungridded visibilites, theoretically allowing for a higher fidelity source subtraction. However, there are a number practical considerations which limit the empirical effectiveness of peeling. Peeling requires an input catalog of sources which are to be peeled. In a real survey, this catalog would be likely be generated from higher frequency observations \citep[eg Parkes Source Catalog,][]{1996yCat.8015....0W}. It would also be possible to be supplement the catalog with data from complementary low frequency facilities (ie GMRT) or, in a bootstrapping fashion, from earlier MWA observations. In any case, the input catalog would invariably contain flux and position errors. Additionally, the observed source positions are perturbed from their true positions by the ionosphere. To account for these errors, the RTS peeling routine uses a local least-squares minimization to fine-tune the position and flux of peeled sources. For our simulations, the input catalogs used by the peeling procedure contain the true (simulated) positions and fluxes of the sources in the field, so that peeling with these values would produce a perfect subtraction. However, the local least squares minimization in this case actually serves to slightly perturb the point source parameters, leading to imperfect subtraction. Further, the ideal method of dealing with diffuse emission when peeling is uncertain. Peeling from the unweighted visibilities would produce flux errors of $\sim 1 \%$ even for the brightest sources. Presently, the MWA RTS implements a simple scheme in which baselines are weighted by a factor of $1 - \mathrm{exp}(- b^2 / {b_0}^2 )$, where $b$ is the baseline length and $b_0$ is a scaling parameter, both in wavelengths. We choose $b_0 = 500$, which effectively saturates the scaling and results in only the longest baselines making a substantial contribution.  Finally, due to the processing time constraints, MWA RTS peeling is non-iterative, meaning that the subtractions of the brightest sources are not updated to account for the subsequently inferred flux of the fainter sources. However, when processing a series of consecutive integrations, the peeling algorithm will make use of previous solutions to improve the current peel. This improvement is not represented in our single snapshot simulations. 

For the ten sources in an empty sky simulations described in Section \ref{10source_emptySky}, the measured relative dynamic range after peeling is $2.2\times 10^3$. With the diffuse sky component added, the measured relative dynamic range is $1.1\times 10^3$. We note that for these bright sources, the performance of our subtraction procedure from the dirty maps is comparable to peeling in the case of no diffuse background ($1.8\times 10^3$ compared $2.2\times 10^3$) and actually performs better in the case of a diffuse background ($1.9\times 10^3$ compared $1.1\times 10^3$). Although these examples do not constitute a definitive comparison of these techniques, they do illustrate that high-fidelity point source subtraction is not necessarily contingent on access to the ungridded visibilities. 

\subsection{A More Realistic Source Distribution}
\label{100sourcesSection}

\begin{figure}[h]
\begin{center}
\includegraphics[scale=0.3, angle=0]{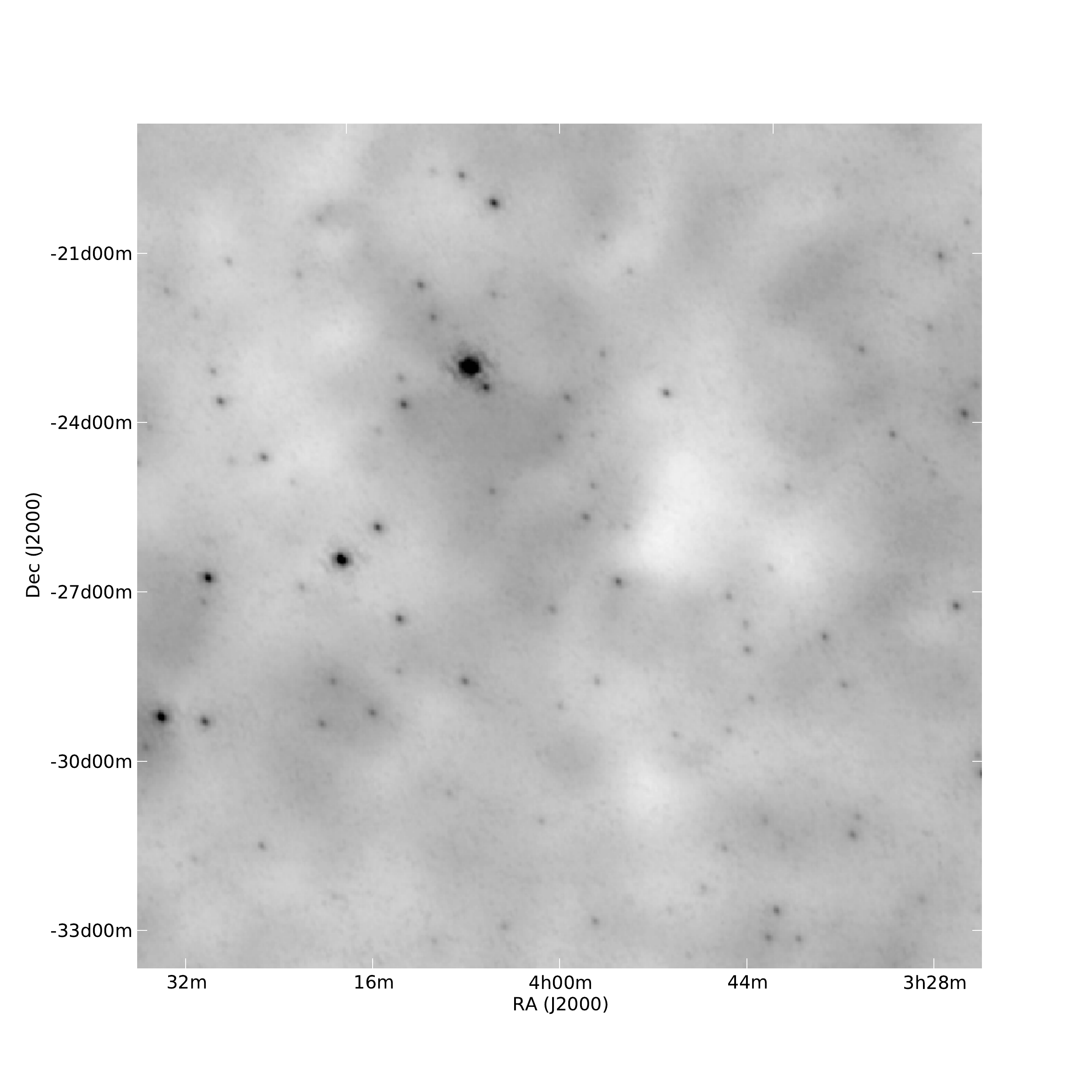}
\caption{100 Point sources together with a diffuse background as described in section \ref{100sourcesSection}}
\label{100SourcesFig}
\end{center}
\end{figure}

As a further illustration of our subtraction procedure, we created a simulations with and without diffuse background together with 
100 sources drawn from a population whose counts are given by

\begin{eqnarray}
N(>S_{Jy}) = N_0 S^{-2.5}_{Jy} Jy^{-1} sr^{-1} 
\end{eqnarray}

\noindent where we restricted the fluxes to be $1-100$ Jy. For comparison, the 6C~151MHz counts \citep{1988MNRAS.234..919H} predict $\sim 200$ source ($> 1 Jy$) across our field of view. The brightest simulated source happens to have a flux of 34.6 Jy. The sources were randomly placed on the sky, subject to a minimum separation of 10 pixels. We found that, unsurprisingly, our perturbative  approach to centroiding produces significant errors for close pairs of sources. For such sources  a simultaneous source fitting procedure such as DAOPHOT \citep{1987PASP...99..191S} may need to be incorporated. Alternately, higher resolution radio data could be used to inform the centroid fits of blended sources. We have also ignored source clustering, although it will be an important consideration in more realistic sky models \citep{2002ApJ...564..576D}. Figure \ref{100SourcesFig} shows the resulting image for the case of sources with a diffuse background. 

\begin{figure}[h]
\begin{center}
\includegraphics[scale=0.4, angle=0]{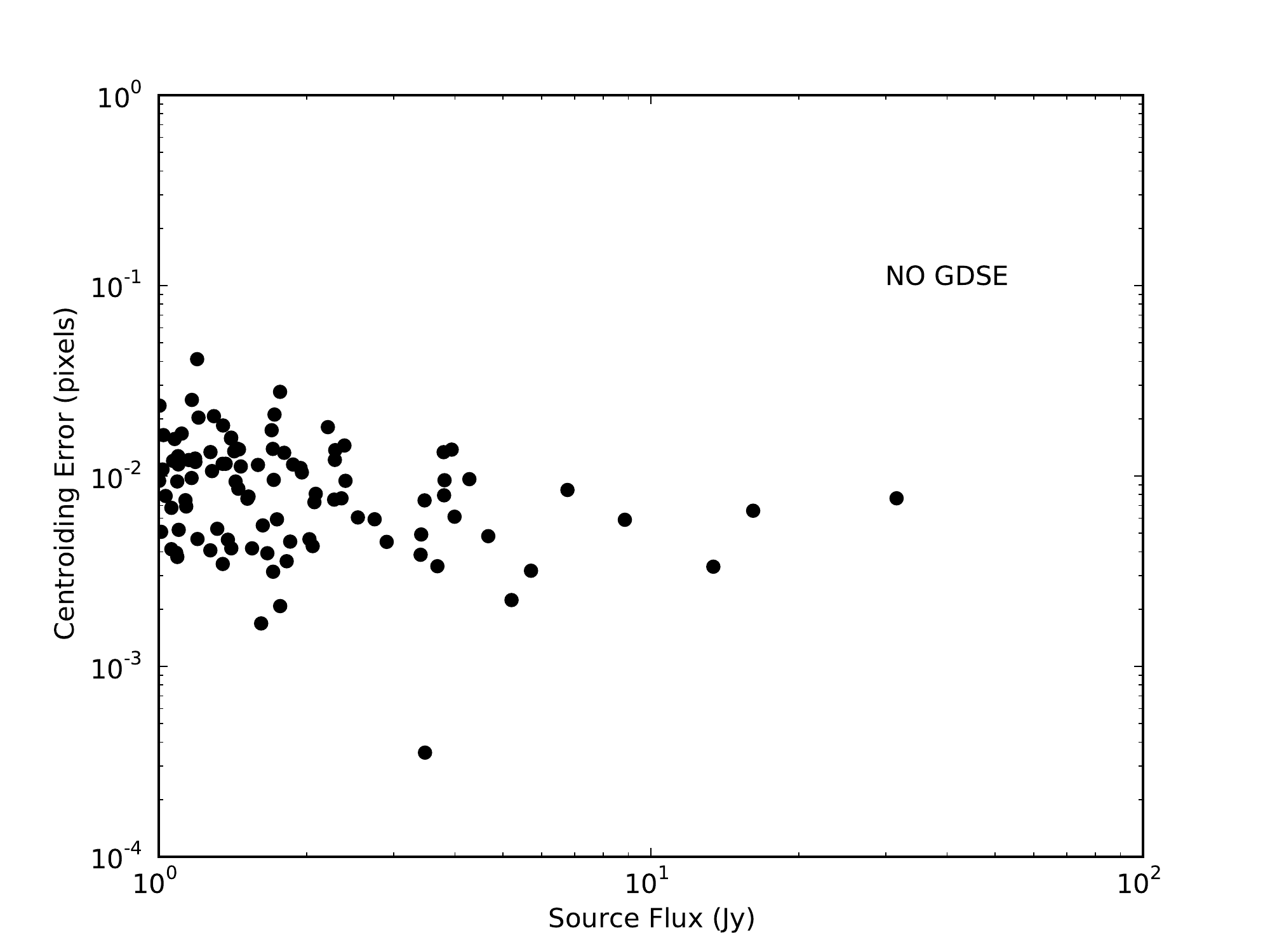}
\caption{The centroiding errors (in pixels) as a function of flux for 100 sources with no diffuse background as described in Section \ref{100sourcesSection}.}
\label{FluxVSCentroidErrorNOGDSE}
\end{center}
\end{figure}

\begin{figure}[h]
\begin{center}
\includegraphics[scale=0.4, angle=0]{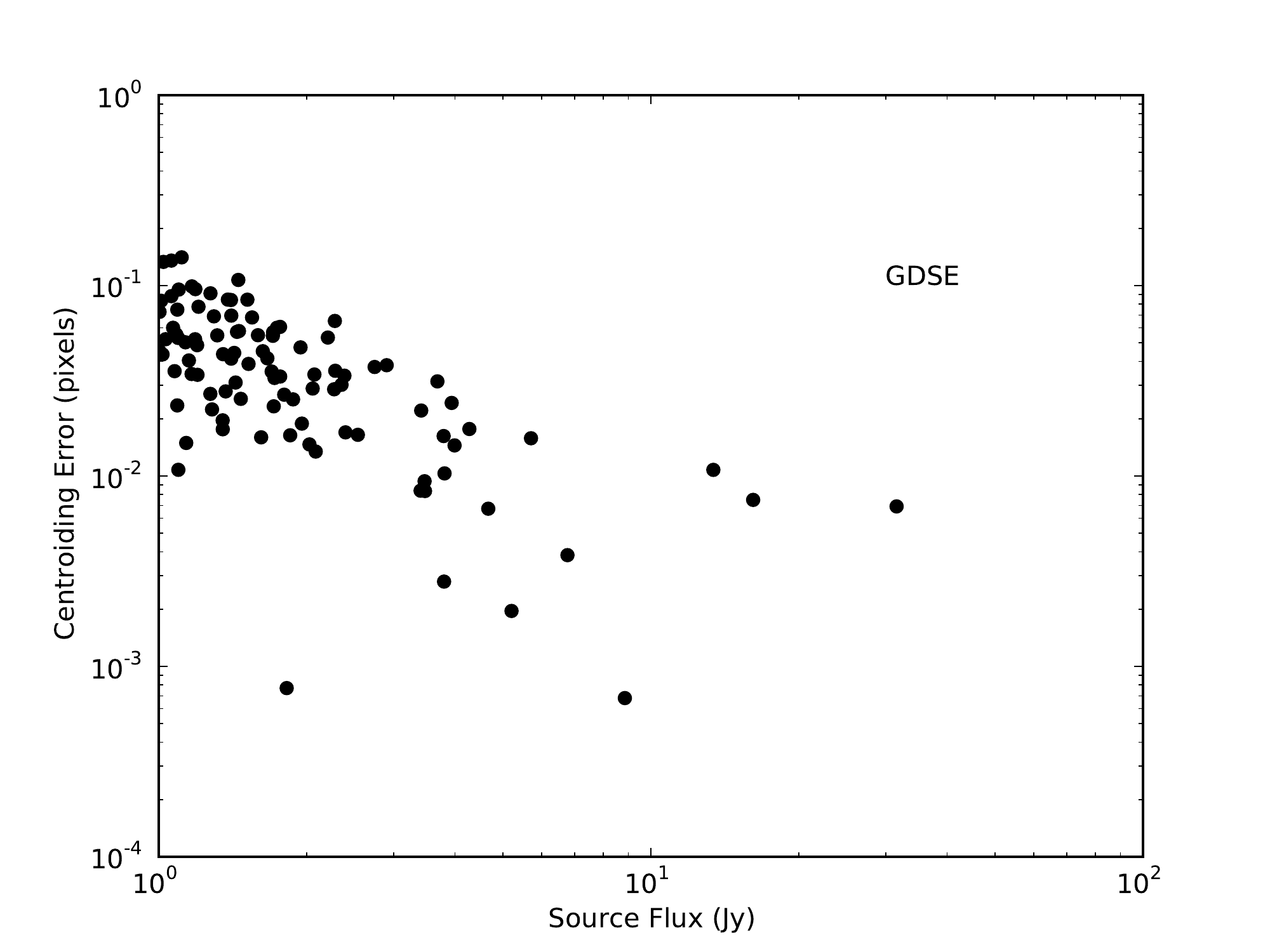}
\caption{The centroiding errors (in pixels) as a function of flux for 100 sources with a diffuse background as described in Section \ref{100sourcesSection}.}
\label{FluxVSCentroidErrorGDSE}
\end{center}
\end{figure}

For the case of 100 sources with no background, the RMS centroiding error following Matched Filtering is $9.7\times 10^{-3}$ pixels and the corresponding improvement in the dynamic range is 190. The PSF Bias Correction is effectively irrelevant due to the magnitude of the input centroid errors. Figure  \ref{FluxVSCentroidErrorNOGDSE} shows the centroiding errors as a function of source flux. The centroiding errors are largely independent of source flux, implying that the sidelobes of the brightest sources are effectively being accounted for while measuring the centroids of the fainter sources. For the case of 100 sources with a diffuse background, the RMS centroiding error following Matched Filtering is 0.04 pixels and the corresponding improvement in the dynamic range is 80. Again, the PSF Bias Correction is effectively irrelevant. The centroiding error is inversely correlated with source flux, as shown in Figure \ref{FluxVSCentroidErrorGDSE}. The centroid fits of the fainter sources are significantly affected by the relatively bright diffuse background. 

\subsubsection{Comparison to EOR Experiment Requirements}

\begin{figure}[h]
\begin{center}
\includegraphics[scale=0.3, angle=0]{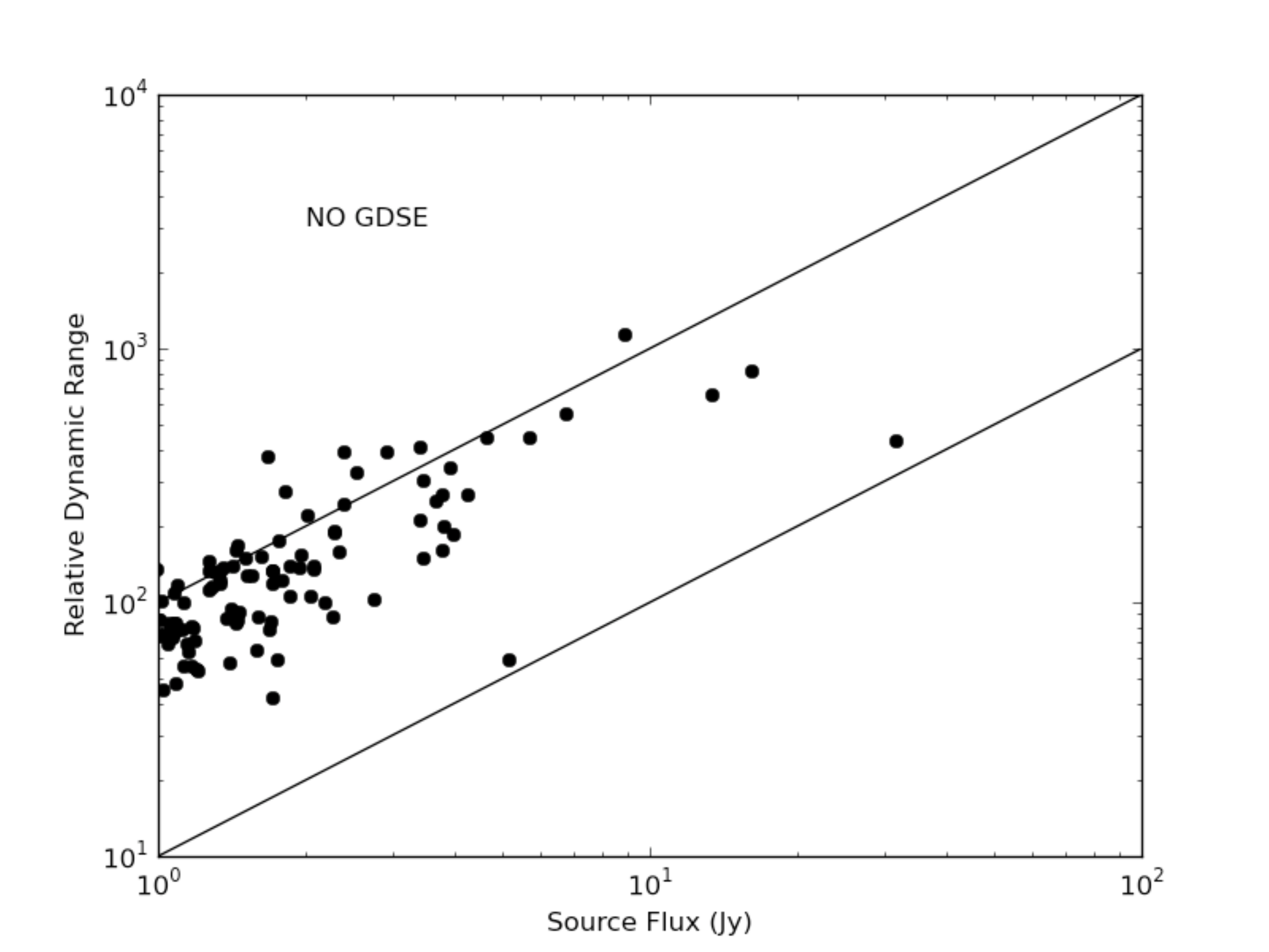}
\caption{The individual relative dynamic range as a function of flux for 100 sources with no diffuse background as described in Section \ref{100sourcesSection}. The two solid lines represent the relative dynamic range corresponding to $S_{cut} = 10$ mJy (upper) and 100 mJy (lower)}
\label{IndividualDRNOGDSE}
\end{center}
\end{figure}

\begin{figure}[h]
\begin{center}
\includegraphics[scale=0.3, angle=0]{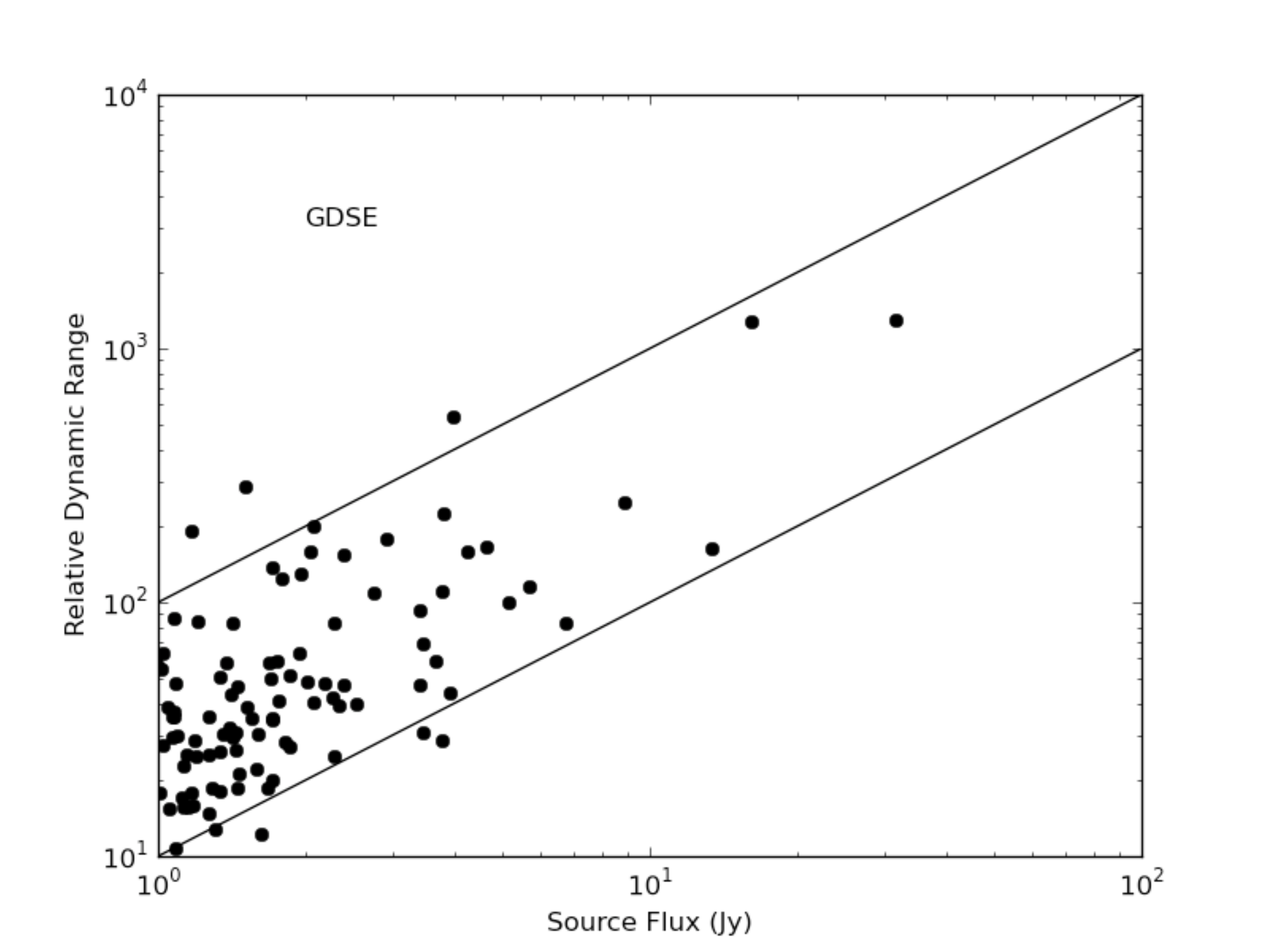}
\caption{The individual relative dynamic range as a function of flux for 100 sources with a diffuse background as described in Section \ref{100sourcesSection}. The two solid lines represent the relative dynamic range corresponding to $S_{cut} = 10$ mJy (upper) and 100 mJy (lower)}
\label{IndividualDRGDSE}
\end{center}
\end{figure}

Since we have access to the true flux and position of each simulated source, we are able to examine the relative dynamic range for each indivdual source. For each source, this quantity is simply the relative dynamic range of the residual image when the selected source has been subtracted with the fitted position and flux, as above, while all of the other sources have been perfectly subtracted using the known true fluxes and positions. In other words, while the above relative dynamic range values measure the factor by which the combined sidelobe level of all of the sources in the image has been reduced, the individual relative dynamic range measures the factor by which the sidelobes of a particular source have been reduced. Figure \ref{IndividualDRNOGDSE} shows the individual relative dynamic range as a function of source flux in the case of no diffuse background. It is evident that the dynamic range is correlated with source flux; the brightest sources are least perturbed by the sidelobes of the other sources and their subtraction is relatively more accurate. This measurement also allows for a rough comparison to the values of $S_{cut}$, the flux limit to which bright sources are required to be removed, as reported by \citet{2009ApJ...695..183B} and \citet{2009MNRAS.398..401L}. For example, suppose that a 1 Jy source was subtracted with a 1\% flux-only error to produce a relative dynamic range of 100. The residuals would then correspond to an unsubtracted 10 mJy source. Such a source is sufficiently faint that the foreground continuum subtraction procedure could tolerate its presence and still achieve the sensitivity required to detect the EOR signal. Although our subtraction procedure produces both flux and centroid errors (as well as beam model errors in the case of calibration errors), the relative dynamic range provides an aggregate measure of the magnitude of the residuals. The two solid lines in Figure \ref{IndividualDRNOGDSE} indicate the relative dynamic range required to correspond to $S_{cut}$ levels of 10 and 100 mJy. Our subtraction exceeds the more optimistic 100 mJy level for all sources and approaches the 10 mJy level for most. Figure \ref{IndividualDRGDSE} shows the individual relative dynamic range as a function of source flux for the case with a diffuse background. As before, the dynamic range is correlated with source flux. In this case, nearly all sources exceed the 100 mJy cut level, but few exceed the 10 mJy level.  

\subsubsection{Comparison with Peeling}

We also applied the RTS peeling, as in Section \ref{peeling_section}, to our 100 source images. For the case of no diffuse background, the relative dynamic range after peeling is $4.6 \times 10^3$. Peeling is more accurate for this example than in the 10 point source case, presumably because the steep source counts imply that the brightest sources can be peeled first with relatively little sidelobe contamination. For the case of 100 sources with a diffuse background, the relative dynamic range after peeling is only 26. The residuals are dominated by a few intrinsically faint sources whose centroid estimates are stochastically degraded by the local least-squares minimization. This example should not be taken as demonstrating the limiting performance of RTS peeling; the peeling results could almost certainly be improved by some algorithmic fine-tuning. However, it does illustrate that a diffuse background will in all likelihood significantly degrade the accuracy of peeling for fainter sources. 

\section{Outstanding Issues}

Our results indicate that subtraction of point sources from dirty maps can substantially improve the sidelobe noise in wide field synthesis images. A simple estimate indicates that the accuracy of our subtraction attains or at least approaches the accuracy required for EOR detection as reported by previous foreground subtraction studies. However, a number of open questions must be resolved before it will be possible to unambiguously determine whether this approach can satisfy the requirements of an EOR experiment. In particular, outstanding issues include:

\begin{itemize}

\item \textbf{Calibration Errors:} The accuracy of the calibration solution will be of vital importance to point source subtraction, both for determining source fluxes and for accurately reconstructing the sidelobe pattern of each source. \citet{2009ApJ...703.1851D} have recently investigated the calibration tolerances for an idealized peeling-like source subtraction. A comprehensive study of the calibration budget will need to account for a combination of complex effects including time-varying calibrators, non-idealized dipole response, and the effects of the ionosphere.
\item \textbf{Multiple Frequencies:} The MWA correlator will simultaneously generate visibilities for hundreds of frequency channels across at 32 MHz bandwidth. Recent work aimed at analyzing CMB data from the Planck satellite \citep{2009MNRAS.394..510H} has demonstrated methods for extending matched filters to point source detection in multi-frequency data. Application of these techniques to MWA data will effectively allow continuum fitting to be combined with the angular power spectrum to increase the contrast between point sources and the diffuse component.  
\item \textbf{Extended Integrations:} As mentioned in Section \ref{simulations}, rotation synthesis will significantly alter (reduce) the sidelobes of all sources. Coadding images produced over an extended observing campaign will also probably require an additional level of regridding. It will also be necessary to model the time dependance of the synthesized beams. Presently, generating simulations which reproduce even a six hour integration is a significant computational challenge. 
\item \textbf{Statistics of Residuals:} We have used the dynamic range as a simple metric for judging the effectiveness of our subtraction procedure. In reality, a more detailed understanding of the subtraction residuals will be required to correctly ascertain the uncertainties in the measured EOR power spectrum. \citet{2009ApJ...695..183B} have suggested the construction of statistical templates which could be used to distinguish foreground residuals from the EOR signal.
\item \textbf{Sky Model:} The sky model of \citep{2008MNRAS.388..247D} uses input data which limit its angular resolution to $\sim 1$ degree. Consequently, spatial power at small angular scales is underrepresented in our diffuse component. A method for introducing additional small scale power into this sky model is currently in development (Bowman, private communication). An alternative is to build up a generic sky model from known sources of emission as done by \citet{2008MNRAS.389.1319J}.
\item \textbf{Out of Beam Sources:} Bright sources which are located outside the primary beam can introduce significant sidelobe noise across the field of view. Our subtraction procedure, which relies upon the ability to locate the main lobe of the synthesized beam, may be inapplicable for such sources. Further, the direction-dependent antenna gains may be poorly constrained towards such sources. A satisfactory approach to dealing with out-of-beam sources remains an open question.

\end{itemize}

\section{Conclusions}

Bright point sources have previously been recognized as an important EOR foreground, but the method by which they should be removed has been unclear. In this work, we have presented a procedure for subtracting point source from radio interferometric synthesis images. We are able to increase the dynamic range of our simulated images by a factor of 2-3 orders of magnitude. The efficacy of this method relies in large part on the excellent $uv$ coverage of the MWA. These results are comparable to the results of the RTS peeling, but are achieved from the dirty maps, alleviating the need for long-term storage of the ungridded visbilities. Of course, peeling is an essential element of the MWA calibration strategy, and hence these two techniques will be used in a complementary fashion; peeling will remove some number of  the brightest sources in real-time, and a subtraction procedure such as we have described will then be used offline to subtract a larger population of fainter sources from the time-averaged dirty maps. Significantly larger image simulations, particularly over longer integrations and multiple observing frequencies, are required to further refine and validate this approach. Nonetheless, our initial estimates indicate that this procedure will be able to remove point sources with sufficient accuracy to satisfy the foreground subtraction requirements of the MWA EOR experiment. 

We thank Gianni Bernardi for helpful comments during the preparation of this work. We acknowledge the support of the Australian Research Council through grants
LE0775621, LE0882938 and DP0877954, the U.S. National Science Foundation through grant AST-0457585 and the Smithsonian Astrophysical Observatory.





\clearpage

\begin{table}[h]
\begin{center}
\caption{Point Source Subtraction Results}\label{centroiding table}
\begin{tabular}{lccc}
\hline
Subtraction Step & Centroid Error$^a$ &  Raw Dynamic Range & Relative Dynamic Range \\ \hline
\hline
\multicolumn{4}{c}{Single Point Source - No Background} \\ \hline 
Dirty Image & 0.02 & $5.3\times10^2$ & 1\\
Matched Filter & 0.002 & $4.2\times10^{5}$ & $7.9\times10^2$ \\
PSF Bias Correction & $10^{-4}$ &  $1.0\times10^8$ & $2.0\times10^4$\\
\hline
\multicolumn{4}{c}{Single Point Source - Diffuse Background} \\

\hline Dirty Image & 0.02 & $5.3\times10^2$ & 1\\
Matched Filter & 5.4$\times10^{-3}$ &  1.6$\times10^5$ & $2.9\times10^2$\\
PSF Bias Correction & $5.5\times10^{-4}$ &  $1.3\times10^6$ & $2.5\times10^3$\\

\hline
\multicolumn{4}{c}{10 Point Sources - No Background} \\

\hline Dirty Image & 0.02 & $1.7\times10^2$ & 1\\
Matched Filter & 1.9$\times 10^{-3}$ &  1.2$\times10^5$ & $6.9\times10^2$\\
PSF Bias Correction & $5.9\times10^{-4}$ &  $3.1\times10^5$ & $1.8\times10^3$\\
RTS Peeling & & & 2.2$\times 10^3$\\
\hline
\multicolumn{4}{c}{10 Point Sources - Diffuse Background} \\

\hline Dirty Image & 0.02 & $1.8\times10^2$ & 1\\
Matched Filter & 9.2$\times10^{-3}$ &  3.4$\times10^4$ & $1.9\times10^2$\\
PSF Bias Correction & $5.5\times10^{-4}$ &  $3.3\times10^5$ & $1.9\times10^3$\\
RTS Peeling & & & 1.1$\times 10^3$\\

\hline
\multicolumn{4}{c}{100 Point Sources - No Background} \\
\hline Dirty Image & 0.02 & $2.4\times10^2$ & 1\\
Matched Filter & 9.7$\times10^{-3}$  &  4.4 $\times 10^3$ & $1.9\times10^2$\\
RTS Peeling & & & 4.6$\times 10^3$\\

\hline
\multicolumn{4}{c}{100 Point Sources - Diffuse Background} \\
\hline Dirty Image & 0.07 & $2.5\times10^2$ & 1\\
Matched Filter & 0.04  &  2.0 $\times 10^3$ & 80\\
RTS Peeling & & & 26\\

\end{tabular}
\medskip\\
$^a$ In pixels. $1$ pixel $= 1.9$ arcmin or $1$ arcsec $\sim 10^{-2}$ pixels. \\
\end{center}
\end{table}

\end{document}